\UseRawInputEncoding

\documentclass[showpacs, oneside, twocolumn, prd, amsmath, amssymb, nofootinbib, superscriptaddress]{revtex4-1}

\usepackage{cases}
\usepackage{amsmath}
\usepackage{amssymb}
\usepackage{amsfonts}
\usepackage{amssymb}
\usepackage{dcolumn}
\usepackage{bm}
\usepackage{bbm}
\usepackage{graphicx}
\usepackage{xcolor}
\usepackage{array}
\usepackage{subfigure}
\usepackage{hyperref}
\usepackage{wasysym}
\usepackage{mathtools}
\usepackage{cleveref}

\def\l{\left}
\def\r{\right}
\def\be{\begin{equation}}
\def\ee{\end{equation}}
\def\ba{\begin{eqnarray}}
\def\ea{\end{eqnarray}}
\def\bl#1\el{\begin{align}#1\end{align}}
\def\nn{\nonumber}
\allowdisplaybreaks[4]

\begin{document}

\title{Induced gravitational waves from statistically anisotropic scalar perturbations}

\author{Chao Chen}
\email{iascchao@ust.hk}
\affiliation{Jockey Club Institute for Advanced Study, The Hong Kong University of Science and Technology, Clear Water Bay, Kowloon, Hong Kong, People's Republic of China}

\author{Atsuhisa Ota}
\email{iasota@ust.hk}
\affiliation{Jockey Club Institute for Advanced Study, The Hong Kong University of Science and Technology, Clear Water Bay, Kowloon, Hong Kong, People's Republic of China}

\begin{abstract}

	Scalar-induced gravitational waves (SIGWs) are attracting growing attention for probing extremely short-scale scalar perturbations via gravitational wave measurements. In this paper, we investigate the SIGWs from statistically anisotropic scalar perturbations, which are motivated in inflationary scenarios in the presence of, e.g., a vector field. While the ensemble average of the SIGW energy spectrum is isotropic for the standard statistically isotropic scalar perturbations, the statistical anisotropy in the source introduces the multipole moments of the differential SIGW energy spectrum. We consider quadrupole anisotropy in the scalar power spectrum and show that the SIGW spectrum has anisotropies up to $\ell=4$. We present generic formulas of the multipole moments and then apply them to the delta-function-like and log-normal source spectra. We find analytic expressions for the former case and show that the infrared scalings of the multipole moments are the same as the isotropic SIGWs. Interestingly, the monopole has an additional local minimum in the high-$k$ tail, a key feature to distinguish from the isotropic SIGWs. The latter log-normal case is analytic for the narrow-peak source, and we perform the numerical calculation for the broad peak. As one expects, the multipole moments become broader with increasing source width. Our results are helpful to test the isotropy of primordial density perturbations at extremely small scales through SIGWs.

\end{abstract}


\maketitle

\section{Introduction}

Gravitational wave (GW) experiments, like LISA \cite{LISA:2017pwj}, and DECIGO \cite{Kawamura:2011zz}, Taiji \cite{Ruan:2018tsw} and TianQin \cite{TianQin:2015yph} will be able to probe the stochastic GW background (SGWB) from astrophysical and cosmological sources. GWs propagate almost freely over space, and they carry information about the Universe much earlier than the recombination epoch that we have already observed via the cosmic microwave background (CMB). Therefore, GW experiments are expected to serve as a promising observational window for unknown physics in the early Universe. Recently, the North American Nanohertz Observatory for Gravitational Waves (NANOGrav) \cite{Arzoumanian:2020vkk} has reported strong evidence of a stochastic common-spectrum process across pulsars from analyzing 12.5-yr pulsar timing array data, which might be
the signal of the SGWB. The SGWB may be interpreted as scalar-induced gravitational waves (SIGWs), which could be a counterpart of primordial black hole (PBH) formation due to large scalar perturbations at extremely short scales \cite{Ashoorioon:2018uey, Ashoorioon:2019xqc, Vaskonen:2020lbd, DeLuca:2020agl, Kohri:2020qqd, Inomata:2020xad, Domenech:2020ers, Cai:2021wzd, Benetti:2021uea}.

In contrast to the usual assumption of the homogeneity and isotropy of the SGWB, the anisotropies can shed light on the unique properties of the source and the propagation over the Universe \cite{Alba:2015cms, Bartolo:2019yeu, ValbusaDallArmi:2020ifo, Bartolo:2019zvb, LISACosmologyWorkingGroup:2022kbp, Dimastrogiovanni:2021mfs}. The first attempt to investigate the anisotropies of the SGWB was made by Ref. \cite{Allen:1996gp} in the case of ground-based interferometers (i.e., LIGO), which has been also considered for space-based interferometers \cite{Cornish:2001hg}, and pulsar timing arrays \cite{Mingarelli:2013dsa, Taylor:2013esa}.
Recently, Ref. \cite{LISACosmologyWorkingGroup:2022kbp} has investigated the sensitivity of LISA to the anisotropies of the SGWB in the millihertz band by using the current instrument specifications, as well as the latest theoretical characterizations of sources of SGWB anisotropies \cite{Contaldi:2016koz, Bartolo:2019oiq, Bartolo:2019yeu, Pitrou:2019rjz}. They found that $\beta \Omega_{\text{GW}} \sim 2 \times 10^{-11}$ (where $\beta$ is the velocity of a boost that induces the dipole) is required to observe a dipole signal with LISA.

In this paper, we investigate another possible origin for anisotropic SIGWs, i.e., the statistical anisotropy of primordial scalar perturbations, which can be realized in, e.g., the anisotropic inflation scenario \cite{Ackerman:2007nb, Soda:2012zm, Maleknejad:2012fw, Bartolo:2011ee}. The statistical properties of the first-order scalar perturbations are transferred to the SIGW energy spectra via second-order coupling in the Einstein equation.  Therefore the observations of SIGWs can be used to probe the statistics of primordial scalar perturbations, which are not accessible via the current observation of the CMB. Previous works on SIGWs mostly assume statistical isotropy of the primordial density perturbations \cite{Ananda:2006af, Baumann:2007zm, Saito:2008jc, Kohri:2018awv, Bartolo:2018rku, Cai:2018dig, Cai:2019jah, Inomata:2020cck, Zhou:2020kkf}.
Statistical anisotropy of scalar perturbations has been considered for induced tensor modes in Ref.~\cite{Ota:2020vfn} for the first time. In that work, the author showed that quadrupole non-Gaussianity introduces superhorizon-induced tensor modes without violating causality at the two-loop level.
Such induced superhorizon tensor modes may be seen in the CMB polarization as in the case with primordial tensor modes.
Recently, Ref. \cite{Dimastrogiovanni:2022eir} explored the enhancement of the propagation anisotropy of SIGWs from the sharply peaked isotropic scalar perturbations.
Our work differs from Refs.~\cite{Ota:2020vfn, Dimastrogiovanni:2022eir}, as we consider the one-loop-order SIGW energy spectrum from statistically anisotropic Gaussian scalar perturbations.
Indeed, evaluation of the one-loop spectrum is more complicated than the soft limit calculation in Ref.~\cite{Ota:2020vfn}, and we report the result in this paper.

The paper is organized as follows: In Sec. \ref{sec:SIGWs}, we first characterize the anisotropies in the differential energy spectra of SIGWs, and then derive the generic expressions of multipole moments of the differential SIGW spectra up to $\ell=4$. Next, we consider the delta-function-like and log-normal spectra for primordial curvature perturbations in Sec. \ref{sec:example} for the radiation-dominated (RD) epoch, and the analytic and numerical expressions for the multipole moments of the differential SIGW spectra are derived. The infrared behaviors and characteristic scales of the multipole moments are discussed as well. Finally, we summarize the results in Sec. \ref{sec:conclusion} and present the details of derivation of the anisotropic spectrum of SIGWs in Appendix \ref{app}.

\section{Induced gravitational waves from anisotropic primordial density perturbations} \label{sec:SIGWs}

The second-order coupling of scalar perturbations in the Einstein equation introduces SIGWs~\cite{Matarrese:1997ay, Noh:2004bc, Wang:2017krj, Wang:2018yql}. Hence, the statistical information of the first-order perturbations can be transferred to SIGWs. In this section, we analyze the effect of the statistical anisotropy of first-order scalar perturbations on the signals of SIGWs.

\subsection{The anisotropic spectrum of SIGWs}

Let us consider SIGWs in the Newtonian gauge, as a SIGW in this gauge is considered physical-i.e., the energy density behaves as radiation in the subhorizon limit \cite{Domenech:2020xin,Ota:2021fdv}. In this paper, we parametrize the metric perturbations as follows:
\be \label{newton_gauge}
ds^2 = a^2(\tau) \Big[ -(1 - 2 \Phi) d\tau^2 + \Big( (1 + 2 \Psi) \delta_{ij} + \frac12 h_{ij} \Big) dx^i dx^j \Big] ~,
\ee
where $\Phi$ and $\Psi$ are the first-order Bardeen potentials, and we define conformal time as
\begin{align}
\tau = \int^t dt a(t) ~.
\end{align}
For simplicity we ignore the linear tensor perturbations, and the second-order tensor perturbation $h_{ij}$ satisfies the transverse-traceless (TT) condition: $\delta^{ik} \partial_k h_{ij} = 0$, $\delta^{ij} h_{ij} = 0$. The latin indices are raised and lowered by Kronecker symbols in this paper.

We consider the commonly used effective energy density of GWs \cite{Brill:1964zz, Isaacson:1968hbi, Isaacson:1968zza, Ford:1977dj, Maggiore:1999vm, Boyle:2005se, Ota:2021fdv}:
\begin{align}
\label{def:rhogw}
\rho_{\text{GW}}(\tau,\mathbf{x}) = { M_{\text{pl}}^2 \over 16 a^2(\tau) } \l\langle h'_{ij}(\tau,\mathbf{x}) h^{ij}{}'(\tau,\mathbf{x}) \r\rangle ~,	
\end{align}
where $M_{\rm pl}\equiv 1/\sqrt{8\pi G}$, a prime is a derivative with respect to conformal time, and the bracket in the first line means the time average over several periods of GWs as well as the ensemble average~\cite{Ford:1977dj}. Equation \eqref{def:rhogw} is justified only for the linear tensor mode in the traditional backreaction formalism, and nonpropagating tensor modes can be included at fourth-order scalar perturbations in general. These are not gravitational waves and cause the gauge dependence issue. Recently, Ref. \cite{Ota:2021fdv} provided a proper interpretation about the gauge transformation of GWs and showed that Eq. \eqref{def:rhogw} can be used to describe the physical SIGW even at fourth order in the scalar perturbations in Newtonian gauge in a general way. Similar discussion has also taken place in Ref.~\cite{Domenech:2020xin}.

A stochastic background of GWs is customarily characterized by their energy density fraction $\Omega_{\text{GW}}$ of the wave vector $\mathbf k$ \cite{Allen:1997ad, Maggiore:1999vm}, which is defined as the GW energy density per unit logarithmic frequency for each line-of-sight direction $\hat{\mathbf{k}}\equiv \mathbf k/k$ namely,
\be \label{omega_GW}
\int_0^\infty {\mathrm{d}k \over k} 
\int {\mathrm{d}{\hat{\mathbf{k}}} \over 4 \pi}
\Omega_{\text{GW}}(\tau, \mathbf{k},\mathbf x) \equiv \frac{\rho_{\text{GW}}(\tau,\mathbf{x})}{\rho_\text{crit}(\tau)} ~,
\ee
where $\rho_\text{crit}(\tau) = 3 M_\text{pl}^2 H^2(\tau)$ with the Hubble parameter $H$.  We assume the statistical homogeneity of the curvature perturbations, so we drop the spatial dependence in $\Omega_{\text{GW}}(\tau, \mathbf k, \mathbf{x})$ in this paper.
Statistical isotropy implies $\hat{\mathbf{k}}$ independence of $\Omega_{\text{GW}}(\tau, \mathbf k)$, so that the angular integral becomes trivial in the standard case.
In this paper, we consider that the SO(3) symmetry of the spectrum is broken to SO(2) in the presence of a preferred direction $\hat{\mathbf{d}}$ in the source. Such a source is generally motivated in inflationary scenarios with spinning fields \cite{Dimastrogiovanni:2010sm,Arkani-Hamed:2015bza,Bartolo:2017sbu,Franciolini:2017ktv}. The anisotropy can be parameterized by the angle between $\hat{\mathbf{d}}$ and $\hat{\mathbf{k}}$. Then, we consider the multipole expansion
\be \label{omega_expansion}
\Omega_{\text{GW}}(\tau, \mathbf{k})
=
\sum_{\ell=0}^{\infty} (-i)^{\ell} (2\ell+1) \Omega_{\ell}(\tau, k) P_{\ell}( \hat{\mathbf{d}} \cdot \hat{ \mathbf{k} } ) ~,
\ee
where $P_{\ell}$ stands for the Legendre polynomials.
We have assumed one preferred direction for simplicity, but multiple preferred directions may be considered.
In that case, we instead consider the expansion with respect to the spherical harmonics but leave the study to the follow-up work.

\medskip
When the relevant modes of GWs are well inside the Hubble radius, one can relate the $\Omega_{\text{GW}}(\tau, \mathbf{k})$ and the power spectrum $\mathcal{P}_h(\tau, \mathbf{k})$ as follows:
\be \label{omega_Ph}
\Omega_{\text{GW}}(\tau, \mathbf{k})
= {1 \over 48} \l( {k \over \mathcal{H}} \r)^2
\overline{ \mathcal{P}_h(\tau, \mathbf{k}) } ~,
\ee
where $\mathcal{H} \equiv a'/a$ is the comoving Hubble parameter, and the overbar denotes the time average over several periods of the GWs. Here, $\mathcal P_h = \sum_{\lambda = +,\times} \mathcal{P}_h^{\lambda\lambda}$, with 
\be \label{ph_hh}
\langle h^\lambda_{\mathbf{k}}(\tau) h^s_{\mathbf{k}'}(\tau) \rangle = \delta^{(3)}(\mathbf{k} + \mathbf{k}') \frac{2 \pi^2}{k^3} \mathcal{P}_h^{\lambda s}(\tau, \mathbf{k}) ~,
\ee
where $\mathcal{P}_h^{\lambda s}(\tau, \mathbf{k})$ is the polarized angular-dependent dimensionless power spectrum for SIGWs.

In the RD epoch, the GW energy spectrum is time independent, and the waves start to decay relative to the matter density after the matter-radiation equality. The energy spectrum observed today $\tau_0$ is given by \cite{Pi:2020otn, Domenech:2021ztg}
\be \label{omega_today}
\begin{aligned}
	&\Omega_{\text{GW}}(\tau_0, f)
	\\\simeq&
	1.6 \times 10^{-5} \l( \Omega_{r,0} h^2 \over 4.18 \times 10^5 \r) \l( g_{*s} \over 106.75 \r)^{-1/3} \Omega_{r,\text{GW}}(\tau_\text{eq}, f) ~,
\end{aligned}
\ee
where $\Omega_{r,\text{GW}}(\tau_\text{eq}, f)$ is the energy spectrum evaluated at the matter-radiation equality $\tau_\text{eq}$, and the physical frequency is related with the comoving scale $k$ as $f= k/(2 \pi a_0) \simeq 1.5 \times 10^{-9} (k / \text{pc}^{-1})$ Hz. $g_{*s}(\tau_\text{ini}) \simeq 106.75$ is the effective degrees
of freedom at the initial production of SIGWs, and $\Omega_{r,0} h^2 \simeq 4.18 \times 10^5$ is the radiation density today given by Planck \cite{Planck:2018vyg}. The relation in Eq.~\eqref{omega_today} will also hold for $\Omega_{\text{GW}}(\tau, \mathbf{k})$. In this paper, we focus on the calculation of the differential energy spectra $\Omega_{\text{GW}}(\tau, \mathbf{k})$ in Eq. \eqref{omega_Ph}, and its current observed spectrum can be directly derived by Eq.~\eqref{omega_today}.

\subsection{Perturbation theory}
In Fourier space, the dynamics of the SIGWs is given by the second-order Einstein equation for the tensor mode
\be \label{eom_hk}
h^{\lambda}_{\mathbf{k}}{}''(\tau) + 2 \mathcal{H} h^{\lambda}_{\mathbf{k}}{}'(\tau) + k^2 h^\lambda_{\mathbf{k}}(\tau) = S^\lambda_{\mathbf{k}}(\tau) ~,
\ee
where the prime denotes the derivative with respect to the conformal time $\tau$.
The source term $S^\lambda_{\mathbf{k}}(\tau)$ is given by \cite{Ananda:2006af, Baumann:2007zm}
\be \label{Sk_bardeen_generic}
\begin{aligned}
S^\lambda_{\mathbf{k}}(\tau)
=&
4 \int \frac{\mathrm{d}^3\mathbf{p}}{(2\pi)^{3/2}} 
\mathbf{e}^\lambda(\mathbf{k},\mathbf{p})
\Big[
2 \Phi_\mathbf{p}(\tau) \Phi_{\mathbf{k} - \mathbf{p}}(\tau)
\\&
+ {4 \over 3 (1 + \omega) } \l( \mathcal{H}^{-1} \Phi_\mathbf{p}'(\tau)+ \Phi_\mathbf{p}(\tau) \r) 
\\& \quad\quad \times
\l( \mathcal{H}^{-1} \Phi_{\mathbf{k} - \mathbf{p}}'(\tau) + \Phi_{\mathbf{k} - \mathbf{p}}(\tau) \r)
\Big] ~,
\end{aligned}
\ee
where $\omega$ is the parameter of the background equation of state-i.e., $\omega = 1/3$ and $0$ for radiation- and matter-dominated epochs, respectively. $\lambda = +, \times$ denote two polarizations of SIGWs. The quantity $\mathbf{e}^\lambda(\mathbf{k},\mathbf{p})$ is defined as $\mathbf{e}^\lambda(\mathbf{k},\mathbf{p}) \equiv e^\lambda_{lm}(\hat{ \mathbf{k} }) p_l p_m $, which is equal to ${1 \over \sqrt{2} } p^2 \sin^2\theta \cos 2\varphi$ for $\lambda=+$ and ${1 \over \sqrt{2} } p^2 \sin^2\theta \sin 2\varphi$ for $\lambda=\times$; where $\cos\theta = \frac{\mathbf{k} \cdot \mathbf{p}}{k p}$ and $(p, \theta, \varphi)$ is the coordinate of $\mathbf{p}$ in a spherical coordinate system whose $(x,y,z)$ axes are aligned with $(e(\hat{ \mathbf{k} }), \bar{e}(\hat{ \mathbf{k} }), \hat{ \mathbf{k} })$; and $(e_i(\hat{ \mathbf{k} }), \bar{e}_i(\hat{ \mathbf{k} }) )$ is a pair of orthogonal polarization vectors, both of which are orthogonal to the wave vector $\mathbf{k}$ of GWs. We assume scalar perturbations are adiabatic for simplicity. Also, we ignore the first-order anisotropic stress, whose effect on SIGWs has been shown to be small \cite{Baumann:2007zm}.

Our calculation is, in principle, similar to the traditional calculations of the isotropic SIGWs \cite{Baumann:2007zm, Ananda:2006af}.
The difference is the angular dependence of the linear scalar power spectrum.
The statistically anisotropic uniform density slice's curvature power spectrum is expanded into
\be \label{expand_pzeta} 
\mathcal{P}_\zeta^{\hat{ \mathbf{d} }}(\mathbf p)
=
\mathcal{P}_\zeta(p) 
\sum_{\ell=0}^\infty 
(-i)^\ell(2\ell+1) A_{\ell}(p) 
P_{\ell}(\hat{\mathbf{d}} \cdot \hat{\mathbf{p}} ) ~,     
\ee
where $\hat{\mathbf{p}} \equiv \mathbf p/p$, and $p\equiv |\mathbf p|$, $\mathcal{P}_\zeta(p)$ is the isotropic part of the dimensionless power spectrum. In a statistically isotropic universe, $A_0=1$ and $A_{\ell\neq 0}=0$. In this paper, we consider nonvanishing $\ell=0$ and $\ell=2$ moments, which are motivated in the anisotropic inflation scenario \cite{Ackerman:2007nb, Soda:2012zm, Maleknejad:2012fw, Dimastrogiovanni:2010sm}, but extension to higher moments are straightforward.  
The first three relevant Legendre polynomials are given as $P_0(\lambda) = 1$, $P_2(\lambda) = (3 \lambda^2 -1)/2$, $P_4(\lambda) = (35 \lambda^4 - 30 \lambda^2 + 3)/8$.

One often uses another convention to parametrize the anisotropy, which can be written as 
\begin{align}
	\mathcal{P}_\zeta^{\hat{ \mathbf{d} }}(\mathbf p)
=\mathcal{P}_\zeta(p)\sum_{\ell=0}^{\infty}g_\ell \left[ 1-(\hat{\mathbf d}\cdot \hat{\mathbf p})^2\right]^\ell ~,\label{gstar}  ~,
\end{align}
and the constraint on the quadrupole moment by the CMB anisotropies is given as $g_2<0.016~(68\%{\rm C.L.})$, which can be recast into $|A_2|<0.0021~(68\%{\rm C.L.})$~\cite{Kim:2013gka}.
However, this limit is applied for the CMB scale $k < 0.1 h/{\rm Mpc}$, which has nothing to do with the scale probed by the SIGWs.
There is no prior reason to simply extrapolate the tight constraints on the statistical isotropy to the unconstrained small scales.
Indeed there is an aniotropic inflation scenario that predicts statistical isotropy at the CMB scale with anisotropic attractor solution at the late stage of inflation~\cite{Soda:2012zm}.

Plugging in the multipole expansion of the scalar power spectrum [Eq. \eqref{expand_pzeta}], we find the power spectrum of SIGWs,
\be \label{ps_gw_full}
\begin{aligned}
	&\mathcal{P}_h^{\lambda s}(\tau, \mathbf{k})
	\\=&
	{k^3 \over \pi}
	\int\mathrm{d}^3\mathbf{p}
	\mathbf{e}^\lambda(\mathbf{k},\mathbf{p})
	\mathbf{e}^s(\mathbf{k},\mathbf{p})
	{ \mathcal{P}_\zeta(p) \mathcal{P}_\zeta(|\mathbf{k} - \mathbf{p}|) \over p^3 |\mathbf{k} - \mathbf{p}|^3 }
	\\& \times
	\sum_{\ell,r=0}^\infty (-i)^{\ell+r} (2\ell+1) (2r+1) 
	A_\ell(p) A_r(|\mathbf{k} - \mathbf{p}|) 
	\\& \times
	P_\ell(\hat{\mathbf d} \cdot \hat{\mathbf{p}} )
	P_r(\hat{\mathbf d} \cdot \widehat{\mathbf{k} - \mathbf{p}} ) F(\tau, p, |\mathbf k-\mathbf p|) ~,
\end{aligned}
\ee
where $\widehat{\mathbf{k} - \mathbf{p}} = ( \mathbf{k} - \mathbf{p} )/ |\mathbf{k} - \mathbf{p}|$, and the source kernel is defined as
\be \label{source_kernel}
F(\tau,p,q)
=
{1\over 2}
\l( {6+6\omega \over 5 + 3 \omega} \r)^4
\l[
\int^\tau_{\tau_\text{ini}} \mathrm{d}\tau_1
g_{k}(\tau,\tau_1)
f(\tau_1, p,q)
\r]^2 ~.
\ee
We also introduce
\be \label{source_fkp}
\begin{aligned}
f(\tau, p, q)
=&
{2 (5 + 3 \omega) \over 3 (1 + \omega) } T(p\tau) T(q\tau)
\\&
+ {2 (1 + 3 \omega) \over 3 (1 + \omega) } \l[ \tau T'(p\tau) T(q\tau) + \tau T(p\tau) T'(q\tau) \r]
\\&
+ {(1 + 3 \omega)^2 \over 3 (1 + \omega) } \tau^2 T'(p\tau) T'(q\tau) ~,
\end{aligned}
\ee
where $\tau_\text{ini}$ is set to zero in this paper, and $g_{k}(\tau,\tau_1)$ is the green function for $h^\lambda_{\mathbf{k}}(\tau)$ in Eq. \eqref{eom_hk}.
$T$ is the linear transfer function for $\Phi$ normalized by superhorizon $\zeta$.
In the present setup, multipole expansion [Eq. \eqref{omega_expansion}] stops at $\ell=4$, since the induced spectrum is given as a product of the scalar power spectrum up to $\ell =2$.
Combining Eqs.~\eqref{omega_Ph} and \eqref{ps_gw_full}, the multipole moments of the SIGW spectrum are written as  
\be \label{ph_expansion}
\Omega_{\ell}(z,k)
=
 {1 \over 48} \l( {k \over \mathcal{H}} \r)^2 H_{\ell}(z, k) ~,
\ee
with $x = |\mathbf{k} - \mathbf{p}| / k$, $y = p/k$, $z = k\tau$, and
\begin{align}
\label{H0}
\begin{split}
	& H_\ell(z, k)
\equiv 
\int_0^{\infty} dy
\int_{|1 - y|}^{1 + y} dx 
\Big[ { 4 y^2 - (1 + y^2 - x^2 )^2 \over 4 x y} \Big]^2
 \\& \times
\overline{F(z, x, y)}
\mathcal{P}_\zeta(ky) \mathcal{P}_\zeta(kx) \big[ \delta_{\ell0} + Q_{\ell}(k,x,y) \big] ~, 
\end{split}
\end{align} 
where we define
\bl \label{Q0}
Q_{0}(k, x,y) \equiv& A_2(kx) A_2(ky) Q_{0xy}(x,y),
\\ \label{Q2}
Q_{2}(k, x,y) \equiv& A_2(kx) Q_{2x}(x,y) + A_2(ky) Q_{2y}(x,y)  \nn
\\& + A_2(kx) A_2(ky) Q_{2xy}(x,y),
\\ \label{Q4}
Q_{4}(k, x,y) \equiv& A_2(kx) A_2(ky) Q_{4xy}(x,y) ~.
\el
The functions $Q_{\ell}(k,x,y)$ contain the information of statistical anisotropy in Eq.~\eqref{expand_pzeta}.
The explicit expressions of $Q_{\ell xy}(x,y)$ are shown in Eqs. \eqref{Q0xy} to \eqref{Q4xy}. 
Thus, $\Omega_2$ and $\Omega_4$ are nonzero for nonvanishing $A_2$.

$H_0$ depends on $A_{2}$, as the product of $P_2$ contains the monopole. This fact is useful in searching for the statistical anisotropy of the primordial curvature perturbations, because the information on the anisotropy can even be extracted from the monopole moment of SIGWs without analyzing the anisotropies.

\section{Examples}  \label{sec:example}

In this section, we evaluate the gravitational wave spectrum for specific examples of scalar power spectra: delta-function and log-normal power spectra.
For simplicity, we assume that $A_\ell$ has no scale dependence at the scale of interest.
We need to calculate the time average $\overline{ F(z, x, y) }$ to obtain $H_{\ell}(z, k)$, which depends on the background evolution of the Universe, i.e., the Green's function in the time integral [Eq. \eqref{source_kernel}], so we need to analyze case by case. The analytic expressions of $\overline{ F(z, x, y) }$ at the RD epoch have been calculated in Refs. \cite{Ananda:2006af, Espinosa:2018eve, Kohri:2018awv}, and Ref. \cite{Kohri:2018awv} also presented the analytic expressions for the matter-dominated epoch. In this paper, we focus on the RD epoch, where $\omega =1/3$ and $\mathcal{H} = 1/\tau$. 

\subsection{Anisotropic SIGWs from a delta-function-like source}

The nearly delta-function-like spectrum of curvature perturbations are realized in several PBH formation models, such as Starobinsky's $R^2$-gravity \cite{Pi:2017gih} and parametric resonance \cite{Cai:2018tuh, Chen:2019zza, Chen:2020uhe}, in which the small-scale curvature perturbations are exponentially amplified over a narrow $k$ region, parametrized as
\be \label{pzeta_delta}
\mathcal{P}_\zeta(k)
= A_\zeta \delta\l( \ln (k/k_*) \r)
= A_\zeta k_* \delta(k - k_*) ~,
\ee
where $k_*$ is the peak position, and $A_\zeta$ is the normalization constant.
While the statistical anisotropies in those scenarios have not been discussed in the literature, we consider the delta function scalar power spectrum as a toy model as the delta function simplifies the convolution integral, so that we can get analytic expressions in this case. 
Equation \eqref{H0} can be recast into
\begin{align}
	\begin{split}
		&H_\ell(z, k) =\int_{- {1 \over \sqrt{2}}}^{{1 \over \sqrt{2}}} \mathrm{d}s 
\int_{{1 \over \sqrt{2}}}^{\infty} \mathrm{d}t
{ (1 - 2 t^2)^2 (1 - 2 s^2)^2 \over 4 (t+s)^2 (t-s)^2 }
\\ &\times 
\mathcal{P}_\zeta\left(k {t-s \over \sqrt{2}} \right) 
\mathcal{P}_\zeta\left(k {t+s \over \sqrt{2}} \right) 
\overline{ F_\text{RD}\left(z, {t-s \over \sqrt{2}}, {t+s \over \sqrt{2}} \right) }
\\
&\times \left[\delta_{0\ell} + Q_{\ell}\left(k,{t-s \over \sqrt{2}},{t+s \over \sqrt{2}} \right) \right] ~,
	\end{split}
\end{align}
where we consider the following quarter turn in the $xy$ plane
\begin{align}
s= {y - x \over \sqrt{2} },~t= {y + x \over \sqrt{2} } ~.
\end{align}

Substituting Eq. \eqref{pzeta_delta} into the multipole moments $H_\ell$, and then using Eq.~\eqref{ph_expansion}, we can calculate the multipole moments of the differential energy spectrum of SIGWs as follows:
\bl \label{omega0_delta}
\Omega_{\ell}^{\delta}(\tilde{k}) 
=&	
\Omega_{\text{iso}}^{\delta}(\tilde{k})
[\delta_{\ell0} + Q_{\ell}^{\delta}(k, 1 / \tilde{k}, 1 / \tilde{k} ) ] ~,
\el
where we introduce
\begin{align}
\begin{split}
	&	\Omega_{\text{iso}}^{\delta}(\tilde{k})  
\equiv  
{3 \over 64} A_\zeta^2
\Big( { \tilde{k}^2 - 4 \over 4 } \Big)^2
\tilde{k}^{2}
(3 \tilde{k}^2 -2)^2
 \\& \times
\Big[
\pi^2 (3 \tilde{k}^2 -2)^2 \Theta( 2/\sqrt{3} - \tilde{k} )
\\&+ \Big(
4 + (3 \tilde{k}^2 -2) \ln\Big| 1 - {4 \over 3 \tilde{k}^2} \Big|
\Big)^2
\Big] \Theta( 2 - \tilde{k} ) ~,	 \label{omegaiso} 
\end{split}
\end{align}
and
\begin{align}
Q_{0}^{\delta}(k, 1 / \tilde{k}, 1 / \tilde{k} )
=&
{5 \over 8} (A_2)^2
( 8 - 12 \tilde{k}^2 + 3 \tilde{k}^4 ) ~,
 \label{Q0_delta}
 \\
\begin{split}
	Q_{2}^{\delta}(k, 1 / \tilde{k}, 1 / \tilde{k} ) 
= & {1 \over 4} A_2 ( 3 \tilde{k}^2 - 4)
 \\
+& {5 \over 56} (A_2)^2 ( 8 + 6 \tilde{k}^2 - 3 \tilde{k}^4 ) ~,
\label{Q2_delta}
\end{split}
\\
Q_{4}^{\delta}(k, 1 / \tilde{k}, 1 / \tilde{k} )
= & {5 \over 448} (A_2)^2 ( 48 + 8 \tilde{k}^2 + 3 \tilde{k}^4 )\label{Q4_delta} ~.
\end{align}
Equation \eqref{omegaiso} is the energy spectrum of SIGWs from an isotropic delta-function-like source \cite{Kohri:2018qtx}. We define the dimensionless wave number $\tilde{k} \equiv k/k_*$. The Heaviside step function $\Theta( 2 - \tilde{k} )$ implies the cutoff at $k = 2 k_*$ which is due to the momentum conservation. The energy spectrum of SIGWs is time independent during RD epoch, which is reasonable, as the short wavelength (i.e., the subhorizon-scale) SIGWs behave like radiation.

\begin{figure*}[ht]
	\centering
	\includegraphics[width=0.45\linewidth]{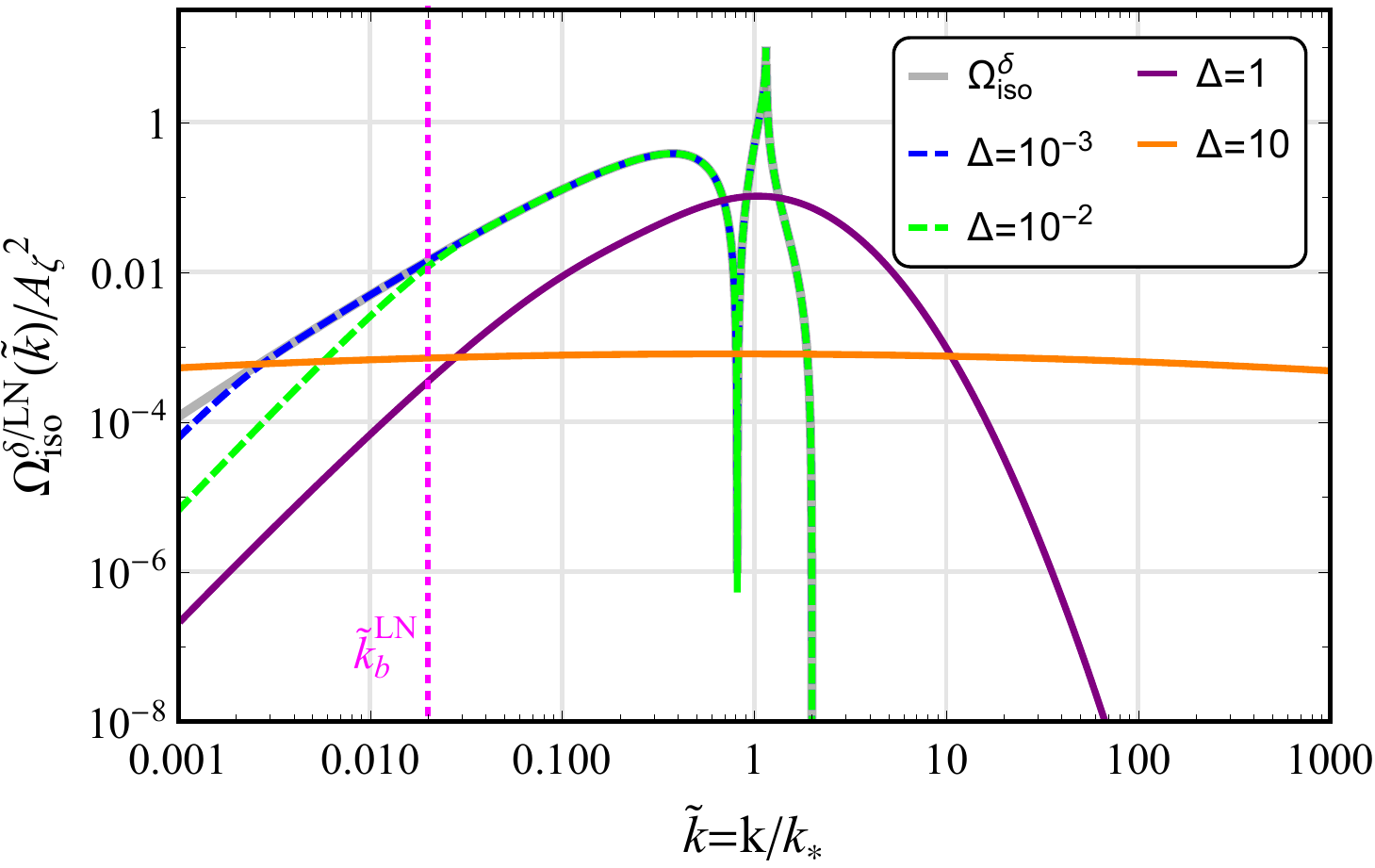}
	\includegraphics[width=0.45\linewidth]{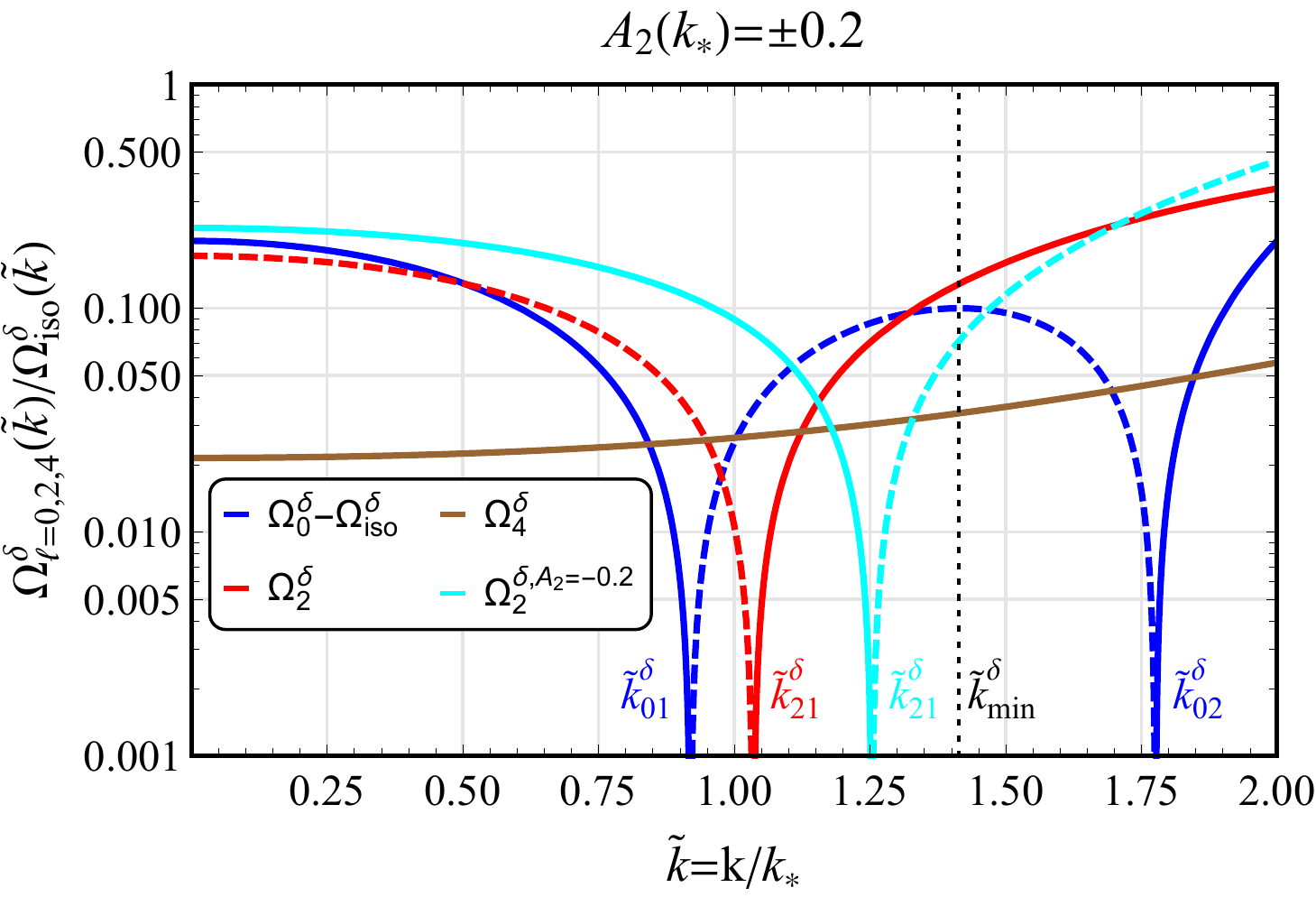}
	\includegraphics[width=0.45\linewidth]{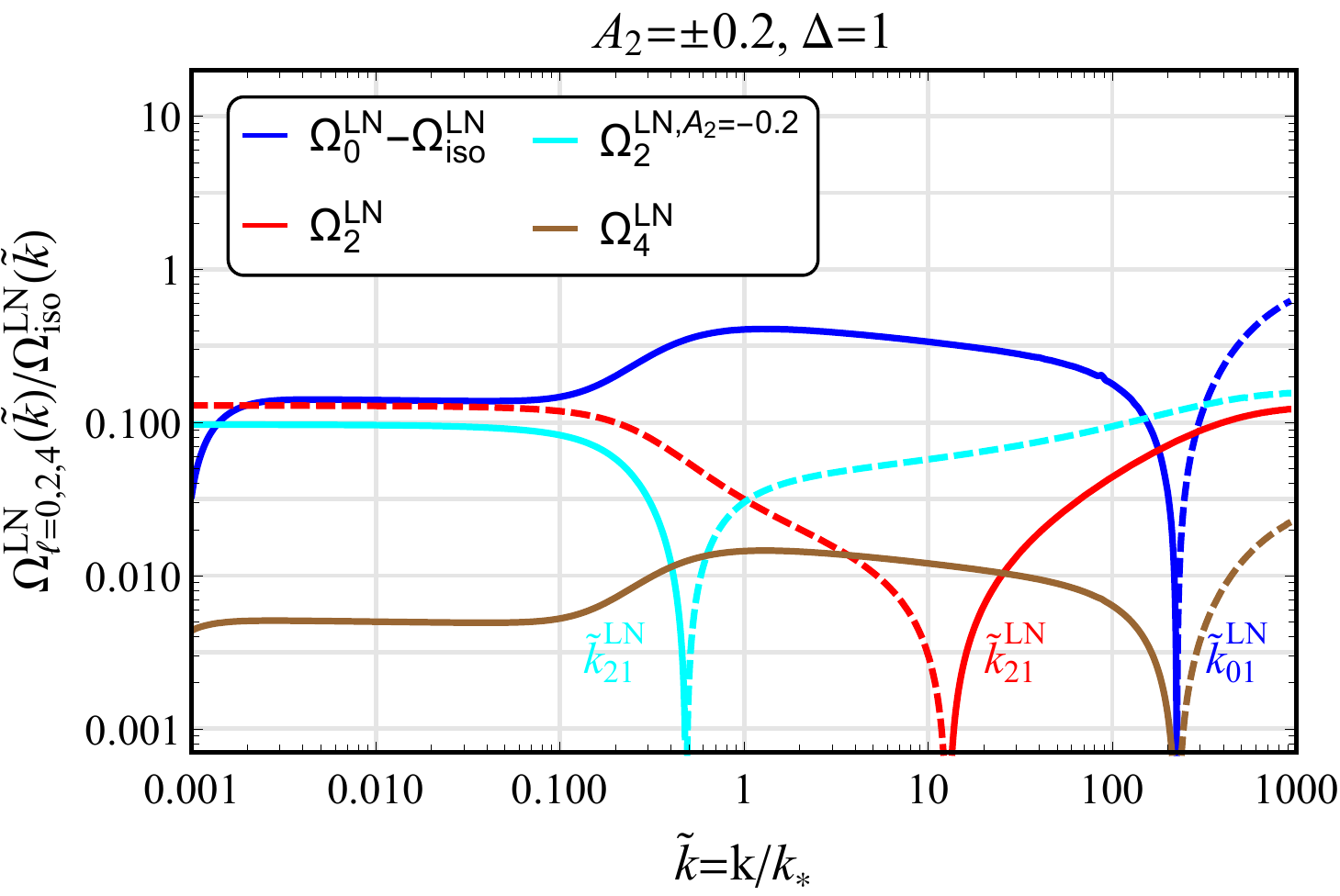}
	\includegraphics[width=0.45\linewidth]{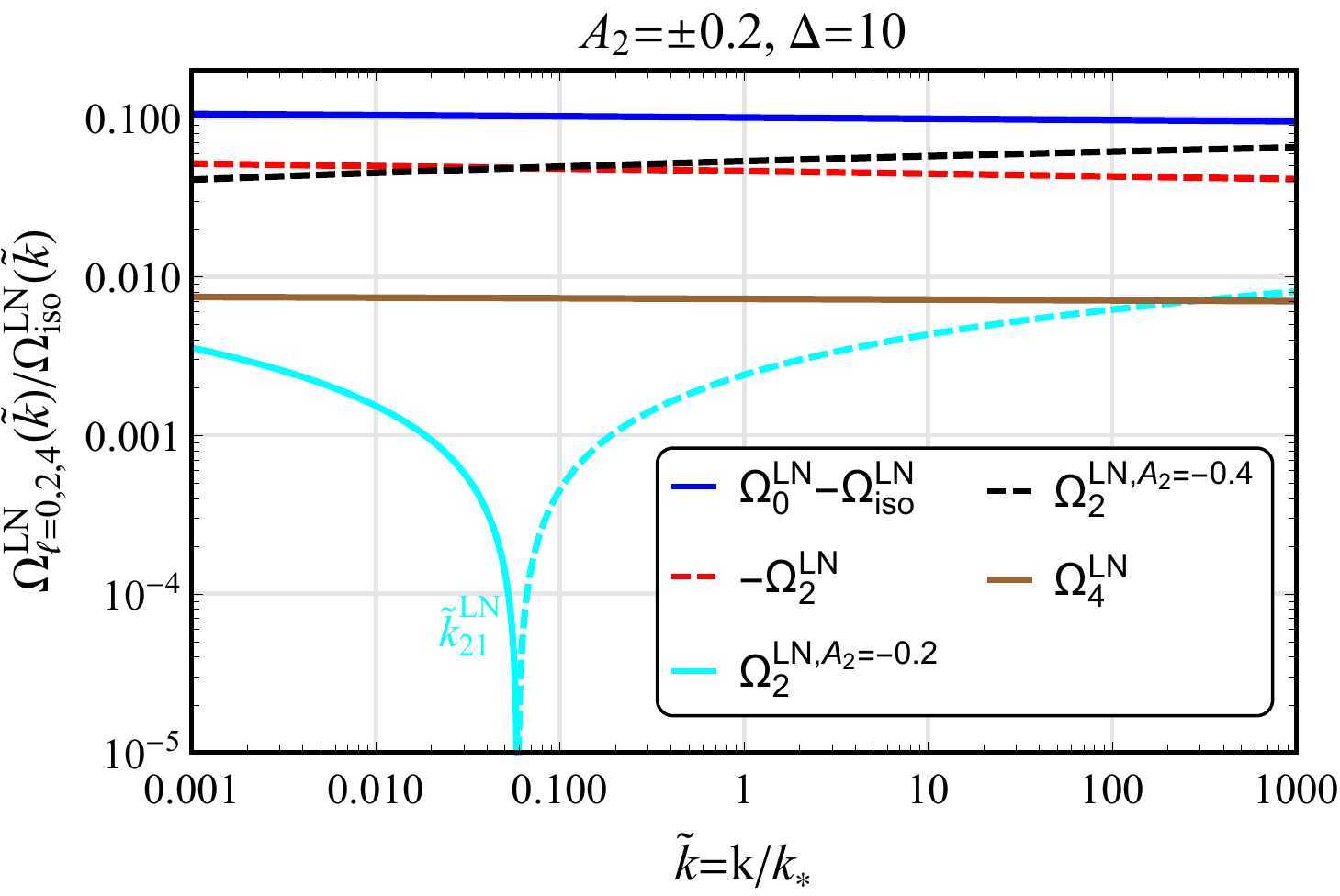}
	\caption{
		{\it Top-left panel}: the isotropic SIGW spectra from the delta-function-like and log-normal sources (with the widths $\Delta = 10^{-3}, 10^{-2}, 1, 10$). The spectra become broader as the width of the source increases. The break scale $\tilde{k}_b^{\text{LN}} = 2 \Delta e^{-\Delta^2}$ for $\Delta=10^{-2}$ is shown as well. 
		{\it Top-right panel}: the relative shapes $\Omega_{\ell}^{\delta}(\tilde{k})/\Omega_{\text{iso}}^{\delta}(\tilde{k})$ as a function of $\tilde{k} = k/k_*$ for $A_2= 0.2$ (blue, red and brown solid curves), and $\Omega_{2}^{\delta}(\tilde{k})/\Omega_{\text{iso}}^{\delta}(\tilde{k})$ for $A_2= -0.2$ (cyan solid curve). The zeros $\tilde{k}_{01}^{\delta}$, $\tilde{k}_{02}^{\delta}$ of $Q_{0}^{\delta}(k, 1 / \tilde{k}, 1 / \tilde{k} )$, and $\tilde{k}_{21}^{\delta}$ of $Q_{2}^{\delta}(k, 1 / \tilde{k}, 1 / \tilde{k} )$ for $A_2= \pm 0.2$ are displayed. The dashed curves denote the absolute value of the negative ratios of  $\ell = 0, 2$, while $Q_{4}^{\delta}(k, 1 / \tilde{k}, 1 / \tilde{k} )$ is always positive. The local minimum of $( \Omega_{0}^{\delta} - \Omega_{\text{iso}}^{\delta} )/\Omega_{\text{iso}}^{\delta}$ located at $\tilde{k}_{\text{min}}^{\delta} = \sqrt{2}$, is labeled by the black dotted line.
		{\it Bottom-left and -right panels}: the numerical results of the SIGW spectra $\Omega_{\ell}^{\text{LN}}(\tilde{k})$ with respect to the isotropic part $\Omega_{\text{iso}}^{\text{LN}}(\tilde{k})$ for the broad peaks $\Delta = 1, 10$, respectively, and the anisotropic coefficients $A_2 = \pm 0.2$. The blue, red, and brown solid curves refer to the monopole, quadrupole and $\ell = 4$ moments of SIGWs, respectively. The blue, red, cyan ($A_2 = - 0.2$) and brown dashed curves refer to the positive expressions $(\Omega_{\text{iso}}^{\text{LN}} - \Omega_{0}^{\text{LN}})/\Omega_{\text{iso}}^{\text{LN}}$, $-\Omega_{2}^{ \text{LN} }/\Omega_{\text{iso}}^{\text{LN}}$ and $-\Omega_{4}^{ \text{LN} }/\Omega_{\text{iso}}^{\text{LN}}$, respectively. The black dashed curve shown in the bottom-right panel refers to $- \Omega_{2}^{ \text{LN} }/\Omega_{\text{iso}}^{\text{LN}}$ with　$A_2 = -0.4$. }
	\label{fig}
\end{figure*}

All effects from the anisotropic source on SIGWs are involved in the relative shapes of $\Omega_{\ell}^{\delta}$ in Eq. \eqref{omega0_delta} with respect to the isotropic SIGWs $\Omega_{\text{iso}}^{\delta}$, which are shown in the top-left panel of Fig. \ref{fig}. The relative shapes of the monopole, quadrupole and $\ell=4$ moments for $A_2 = 0.2$ are displayed in the top-right panel of Fig. \ref{fig}, by blue, red and brown solid curves, respectively. The cyan curve denotes the quadrupole for $A_2 = -0.2$, while $\Omega_{0}^{\delta}$ and $\Omega_{4}^{\delta}$ are invariant under the transformation $A_2 \rightarrow - A_2$. This distinct behavior of the quadrupole moment is due to the linear term in terms of $A_2$, stemming from the coupling between the monopole and quadrupole moments of source.
From Eqs. \eqref{Q0_delta}-\eqref{Q4_delta}, we have $\Omega_{0}^{\delta}(\tilde{k}) \propto 1 + (A_2)^2$, $\Omega_{2}^{\delta}(\tilde{k}) \propto A_2$ and $\Omega_{4}^{\delta}(\tilde{k}) \propto (A_2)^2$.

It is straightforward to see from Eqs. \eqref{Q0_delta}-\eqref{Q4_delta} that, while $Q_{4}^{\delta}$ is always non-negative, $Q_{0,2}^{\delta}$ can be either positive or negative. The zeros for $Q_{0,2}^{\delta}$ are labeled as $\tilde{k}_{\ell n}^{\delta}$, which depends on the size of $A_2$: $\tilde{k}_{0 n}^{\delta} = \sqrt{2 \pm 2/\sqrt{3}}$ and $\tilde{k}_{21}^{\delta} \in [1.04, 1.33]$ for $-0.4 \leq A_2\leq 0.2$ (which is required by the positivity of the scalar power spectrum in Eq. \eqref{expand_pzeta}. We emphasize that this $A_2$ constraint merely comes from the truncation of the multipole expansion in Eq.~\eqref{expand_pzeta} at quadratic order, which is the case in the anisotropic inflationary model we considered. In general, inflationary models provide multipole coefficients such that the scalar power spectrum is non-negative definite, and it is not necessarily possible to have a simple constraint equations for each coefficient.
For the monopole moment $\Omega_{0}^{\delta}$, there is an extra narrow dip around the scale $\tilde{k}_{\text{min}}^{\delta} = \sqrt{2}$ shown in the top-right panel of Fig. \ref{fig}. The width of this dip is estimated as
\be \label{dip_delta}
\tilde{k}_{02}^{\delta} - \tilde{k}_{01}^{\delta} \simeq 0.86 ~,
\ee
and the local minimum is given by
\be \label{min_omega_delta}
\Omega_{0}^{\delta}(\tilde{k}_{\text{min}}^{\delta}) 
=
\Omega_{\text{iso}}^{\delta}(\tilde{k}_{\text{min}}^{\delta})
\l[ 1 - {5 \over 2} (A_2)^2 \r] ~.
\ee
This unique feature in principle can be used to extract the magnitude of $A_2$ when compared with the isotropic case. When we consider the positivity constraint $- 0.4 \leq A_2 \leq 0.2$, there will be at most $40\%$ deviation from the isotropic SIGWs at $\tilde{k}_{\text{min}}^{\delta}$, which is one of the main results in this paper. 
In addition, the small amplifications occur in the infrared regime and high-$k$ tail of $\Omega_{0}^{\delta}$; however, the total amplitudes are suppressed in these ranges. For the quadrupole moment $\Omega_{2}^{\delta}$, the zero point $\tilde{k}_{21}^{\delta}$ determines the positive or negative contribution of $\Omega_{2}^{\delta}$ to the differential energy spectrum $\Omega_{\text{GW}}(\tau, \mathbf{k})$ in Eq. \eqref{omega_expansion} for various $k$ ranges.

The infrared behavior of the energy spectrum is critical for GW observations \cite{Cai:2019cdl, Domenech:2020kqm, Pi:2020otn}. It has already been shown in Ref. \cite{Cai:2019cdl} that there in general exists an universal infrared behavior $k^3$ of SIGWs no matter the super- or subhorizon scales when several physical conditions are satisfied. However, the isotropic SIGWs from the delta-function-like source are shown to have $k^2$ infrared scaling when we take the limit $\tilde{k} \ll 1$ and yield $\Omega_{\text{iso}}^{\delta}(\tilde{k}) \simeq 3 A_{\zeta}^2 \tilde{k}^2 \ln^2\tilde{k}$. In this case the delta-function-like source is unphysical, as the delta function in Fourier space implies a two-point correlation for infinitely long distance in real space.
From Eqs. \eqref{Q0_delta}-\eqref{Q4_delta}, it is straightforward to see that the functions $Q_{\ell}^{\delta}(k, 1/\tilde{k},1/\tilde{k})$ are constant for $\tilde{k} \ll 1$, so the multipole moments $\Omega_{\ell}^{\delta}(\tilde{k})$ have the same infrared scaling as the isotropic SIGWs $\Omega_{\text{iso}}^{\delta}(\tilde{k})$, which can be clearly seen in Fig. \ref{fig}. This conclusion can also be immediately seen from Eq. \eqref{ps_gw_full} when we take the infrared limit $k \ll 1$, the direction dependence of $\mathbf{k}$ vanishes as $\widehat{\mathbf{k} - \mathbf{p}}\to -\hat {\mathbf p}$ in that limit; therefore, the infrared behavior must be same with the isotropic case. Hence, we conclude that it is hard to distinguish the SIGWs from the anisotropic part with the isotropic SIGWs in the infrared regime.

\subsection{Anisotropic SIGWs from a log-normal source}

The delta-function-like spectrum is an unphysical toy model to approximate a sharp peak in the power spectrum. Realistic peaks could be approximated by the log-normal spectrum with a nonzero peak width. 
To our knowledge, the isotropic SIGWs from a log-normal source are first calculated in Ref. \cite{Pi:2020otn}; we here mainly follow their treatments therein. The log-normal spectrum is parametrized as
\be \label{lognormal_pzeta}
\mathcal{P}_\zeta(k)
= { A_\zeta \over \sqrt{2 \pi} \Delta} \exp\Big[ - { \ln^2(k/k_*) \over 2 \Delta^2} \Big] ~,
\ee
where $ A_\zeta = \int_{-\infty}^{\infty} \mathcal{P}_\zeta(k) \mathrm{d} \ln k$ is the normalization constant, $\Delta$ is the variance describing the width of $\mathcal{P}_\zeta(k)$ and $\ln k_*$ is the mean location of this log-normal distribution. We note that
the log-normal spectrum [Eq. \eqref{lognormal_pzeta}] reduces to the monochromatic spectrum [Eq. \eqref{pzeta_delta}] in the small-width limit. On the other hand, if the log-normal distribution is broad enough, $\Delta \rightarrow \infty$, but keeping the ratio $A_\zeta / \Delta$ fixed, it would recover the scale-invariant power spectrum which is favored by CMB observation on the large scales.
The previously studied statistically anisotropic scalar spectrum in anisotropic inflation could also be classified into the latter category.

The multipole moments of the energy spectrum of SIGWs originated from the log-normal source [Eq. \eqref{lognormal_pzeta}] can also be calculated by using the generic formula \eqref{H0}, and we get
\begin{align}
\begin{aligned}
	\label{H0_logxy}
	&H_\ell^{\text{LN}}(z,x,y)
	\\ =&
	{ A_\zeta^2 \over 2 \pi \Delta^2 }
	\int_0^{\infty} dy
	\int_{|1 - y|}^{1 + y} dx 
	\Big[ { 4 y^2 - (1 + y^2 - x^2 )^2 \over 4 x y} \Big]^2
	\\& \times
	\exp\Big[ - { \ln^2x + \ln^2y + 2 \ln\tilde{k} \ln(xy) + 2 \ln^2\tilde{k} \over 2 \Delta^2} \Big]
	\\& \times
	\overline{F(z, x, y)} \big[ \delta_{\ell 0} + Q_{\ell}(k,x,y) \big] ~.	
\end{aligned}
\end{align}
Reference \cite{Pi:2020otn} found the following convenient coordinate transformation: 
\begin{align}
u = {1\over\sqrt{2}} \ln(xy),~v = {1\over\sqrt{2}} \ln{x \over y} ~.
\end{align}
In the new frame, the integral domain is enclosed by the following curves:
\begin{align}
\chi(u) &= \sqrt{2} \text{arccosh}\l( e^{- u / \sqrt{2} } /2 \r) ~,
\\
\xi(u) &= \sqrt{2} \text{arcsinh}\l( e^{- u/\sqrt{2} } /2 \r) ~. 	
\end{align}
Using the above new variables, Eq.~\eqref{H0_logxy} becomes
\be \label{H0_logst}
\begin{split}
&H_\ell^\text{LN}(z, k)
=
{729 \over 16} \l( { 6+6\omega \over 5 + 3 \omega } \r)^4 { A_\zeta^2 \over \pi \Delta^2 z^2}\tilde{k}^2
e^{\Delta^2}
\\
&\times 
\int_{-\infty}^{\infty} \mathrm{d}u
\int_{\text{Re}[\chi(u)] }^{\xi(u)} \mathrm{d}vR\left(e^{\frac{u+v}{\sqrt2}},e^{\frac{u-v}{\sqrt 2}}\right)
\\
&\times 
\exp\l[ - { v^2 \over 2 \Delta^2} \r] 
\exp\l[ - { \big( u + \sqrt{2} (\ln\tilde{k} + \Delta^2 ) \big)^2 \over 2 \Delta^2} \r]
\\
&\times 
 \left[\delta_{\ell 0}+ {Q}_{\ell} \left(k,e^{\frac{u+v}{\sqrt2}},e^{\frac{u-v}{\sqrt 2}}\right) \right] ~,	
\end{split}
\ee
where we introduce
\begin{align}
\begin{split}
\label{Rlogst}
	&R(x,y)
	\\ \equiv &
	{\l(x^2 + y^2 - 3 \r)^4 \l[x^4 + (y^2 - 1)^2 - 2 x^2 (y^2 + 1) \r]^2 \over 1024 x^6 y^6}
	\\& \times
	\Bigg\{ \pi^2 \Theta\l[ 2 \sqrt{x y} \cosh\l( {1 \over 2} \ln{x \over y} \r)  - \sqrt{3} \r]
	\\& \quad
	+ \l( \ln\l| { (x+y)^2 - 3 \over (x-y)^2 - 3 } \r| - { 4 x y \over x^2 + y^2 - 3 } \r)^2
	\Bigg\} ~.	
\end{split}
\end{align}
Note that the integrands of $H_{\ell}^\text{LN}(z, k)$ in Eq.~\eqref{H0_logst} are even in $v$, and we only need to calculate the integrals over the upper-half domain, which is formed by $\text{Re}[\chi(u)]$ and $\xi(u)$; the real part in $\chi(u)$ is taken to ensure that $\chi(u) = 0$ for $u > - \sqrt{2} \ln2$. Since the integrands of $H_{\ell}^\text{LN}(z, k)$ are proportional to the Gaussian function, the result depends crucially on whether the widths of the Gaussian peaks are inside the integration domain or not, which are determined by the values of $\Delta$. Following Ref. \cite{Pi:2020otn}, we will discuss the cases of a narrow peak $\Delta \ll 1$ and a wide peak $\Delta \gtrsim 1$ separately.

\subsubsection{Narrow peak}

The SIGW spectrum for a narrow peak $\Delta \ll 1$ has a similar form as the delta-function-like power spectrum.
The main contribution of integrals over $u$ and $v$ comes from the peaks of Gaussian functions-i.e., $u = - \sqrt{2} (\ln\tilde{k} + \Delta^2 )$ and $v=0$.
Using the method of stationary phase, we approximately perform the $t$ integral as follows:
\begin{align}
\begin{split}
	&\int_{\text{Re}[\chi(u)] }^{\xi(u)} \mathrm{d}v
\exp\l[ - { v^2 \over 2 \Delta^2} \r]\\
&\times
{Q}_{\ell}\left(k,e^{\frac{u+v}{\sqrt2}},e^{\frac{u-v}{\sqrt 2}}\right)
R\left(e^{\frac{u+v}{\sqrt2}},e^{\frac{u-v}{\sqrt 2}}\right)
\\\simeq&
\sqrt{ {\pi \over 2} }
{Q}_{\ell}\left(k,e^{\frac{u}{\sqrt2}},e^{\frac{u}{\sqrt 2}}\right)
R\left(e^{\frac{u}{\sqrt2}},e^{\frac{u}{\sqrt 2}}\right) 
\\
&\times \Delta
\l[ \text{erf}\l( { \xi(u) \over \sqrt{2} \Delta} \r)
- \text{erf}\l( { \text{Re}[\chi(u)] \over \sqrt{2} \Delta} \r) \r] ~,
\end{split}	
\end{align}
where the error function is defined as 
\begin{align}
	\text{erf}(w) \equiv {2 \over \sqrt{\pi} } \int_{ 0 }^{w} e^{- z^2} dz ~.	
\end{align}
Substituting the above expressions back into Eq. \eqref{H0_logst}, we similarly integrate $v$ and get
\bl \label{omega0_log}
\Omega_{0}^{\text{LN}}(\tilde{k}) 
=&	
\Omega_{\text{iso}}^{\text{LN}}(\tilde{k})
\big[ 1 + {Q}_{0}^{\delta}\big(k,1/(e^{\Delta^2}\tilde k),1/(e^{\Delta^2}\tilde k)\big) \big] ~,
\\ \label{omega2_log}
\Omega_{2}^{\text{LN}}(\tilde{k})  
=&
\Omega_{\text{iso}}^{\text{LN}}(\tilde{k})
{Q}_{2}^\delta\big(k,1/(e^{\Delta^2}\tilde k),1/(e^{\Delta^2}\tilde k) \big) ~,
\\ \label{omega4_log}
\Omega_{4}^{\text{LN}}(\tilde{k}) 
=&
\Omega_{\text{iso}}^{\text{LN}}(\tilde{k})
{Q}_{4}^\delta\big(k,1/(e^{\Delta^2}\tilde k),1/(e^{\Delta^2}\tilde k) \big) ~,
\el
where $\Omega_{\text{iso}}^{\text{LN}}(\tilde{k})$ is the energy spectrum for the isotropic SIGWs \cite{Pi:2020otn},
\begin{align}
	\label{omega_iso_log}
\begin{split}
&\Omega_{\text{iso}}^{\text{LN}}(\tilde{k})=\frac{\Omega_{\rm iso}^{\delta}\left(e^{\Delta^2}\tilde k\right)}{4}
\l[ \text{erf}\l( { 1 \over \Delta} \text{arcsinh}{ \tilde{k} e^{\Delta^2} \over 2 } \r)\r.
\\
&\l. 
- \text{erf}\l( { 1 \over \Delta} \text{Re}\l( \text{arccosh}{ \tilde{k} e^{\Delta^2} \over 2 } \r) \r) \r]	~.
\end{split}
\end{align}
Here we take $\omega = 1/3$ for the radiation domination.
It is straightforward to see that the above expressions recover the delta-function-like case, Eqs. \eqref{Q0_delta}-\eqref{Q4_delta}, when the width of peak vanishes, $\Delta \rightarrow 0$. As we expect, the SIGWs from a log-normal source reduce to the delta-function-like case-i.e.,
\begin{align}
\lim_{\Delta\to 0}\Omega_{\ell}^{\text{LN}}(\tilde{k}) = \Omega_{\ell}^{\delta}(\tilde{k}) ~.	
\end{align}
The narrow-peak result of the isotropic SIGWs [Eq. \eqref{omega_iso_log}] can be further simplified by using the approximation $e^{\Delta^2} \simeq 1$ \cite{Pi:2020otn}: 
\begin{align}
\Omega_{\text{iso}, \Delta \ll 1}^{\text{LN}}(\tilde{k}) \simeq \text{erf}\l( { 1 \over \Delta} \text{arcsinh}{ \tilde{k} \over 2 } \r) \Omega_{\text{iso}}^{\delta}(\tilde{k}) ~.	
\end{align}
Note that the error function is independent of the kernel of the source, so it is also independent of the background equation of state \cite{Pi:2020otn}. Similarly to the delta-function-like case, there are also corresponding zero points of $Q_{0,2}^{\text{LN}}$: 
\be \label{narr_broad_relation}
\begin{aligned}
\{ \tilde{k}_{01}^{\text{LN}}, \tilde{k}_{02}^{\text{LN}}, \tilde{k}_{21}^{\text{LN}} \}
= 
e^{- \Delta^2} \{ \tilde{k}_{01}^{\delta}, \tilde{k}_{02}^{\delta}, \tilde{k}_{21}^{\delta} \} ~.
\end{aligned}
\ee
These characteristic scales for the narrow peak are related to those of the delta-function-like case by a factor $e^{- \Delta^2}$, which is a universal corresponding relation between the narrow-peak and delta-function-like cases. 
Note that the coefficient $A_2$ is assumed to be a constant in this paper, so these zero points $\tilde{k}_{\ell n}^{\text{LN}}(A_2)$ depend on the width $\Delta$ only, we recover the delta-function-like case when the narrow limit is taken-i.e., $\tilde{k}_{\ell n}^{\text{LN}} \simeq \tilde{k}_{\ell n}^{\delta}$ as $\Delta \rightarrow 0$. Similarly, the monopole moments provide the major contribution for the small $A_2$.

A distinctive feature of the narrow log-normal case compared to the delta-function-like case is the infrared behavior of SIGWs' energy spectra. Reference \cite{Pi:2020otn} shows that there exists a break scale $\tilde{k}_b^{\text{LN}} = 2 \Delta e^{-\Delta^2}$ for a narrow log-normal source, where the GW spectrum changes its infrared behaviors from $k^3$ to $k^2$.
The infrared scaling $k^3$ on the superhorizon scales originates from the causality \cite{Liddle:1999hq, Cai:2019cdl}, while the delta-function-like source merely gives $k^2$ infrared scaling.
This is because the delta-function-like source [Eq. \eqref{pzeta_delta}] is not physical as we include the infinite-distance correlations in the real space. In addition, there is also a logarithmic divergence at $\tilde{k}_{p}^{\text{LN}} = 2/\sqrt{3} e^{-\Delta^2}$, a local infrared maximum at $\tilde{k}_{\text{IR}}^{\text{LN}} = e^{- 1 -\Delta^2}$ and a dip at $\tilde{k}_{d}^{\text{LN}} = \sqrt{{2\over3}} e^{-\Delta^2}$.
Equations \eqref{omega0_log}-\eqref{omega4_log} imply that the infrared behaviors of $\Omega_{\ell}^{\text{LN}}(\tilde{k})$ are determined by the isotropic part $\Omega_{\text{iso}, \text{IR}}^{\text{LN}}(\tilde{k})$. 
Similar to the delta-function-like case, there also exists a local minimum at $\tilde{k}_{\text{min}}^{\text{LN}} = e^{- \Delta^2} \tilde{k}_{\text{min}}^{\delta}$ for the monopole moment $\Omega_{0}^{\text{LN}}(\tilde{k})$. The width of the dip around $\tilde{k}_{\text{min}}^{\text{LN}}$ is estimated as
\be \label{dip_LN}
\tilde{k}_{02}^{ \text{LN} } - \tilde{k}_{01}^{ \text{LN} }
= e^{- \Delta^2} (\tilde{k}_{02}^{\delta} - \tilde{k}_{01}^{\delta}) ~,
\ee
and the local minimum is given by
\be
\Omega_{0}^{ \text{LN} }(\tilde{k}_{ \text{min}}^{\text{LN}})
=
\Omega_{\text{iso}}^{\text{LN}}(\tilde{k}_{\text{min}}^{\text{LN}})
\l[ 1 - {5 \over 2} (A_2)^2 \r] ~,
\ee
which is also at most a $40\%$ deviation from the isotropic SIGWs.

\subsubsection{Broad peak}

A variety of models predict a broad primordial curvature spectrum-e.g., Refs \cite{Garcia-Bellido:2017mdw, Cheng:2018yyr, Kohri:2012yw, Ando:2017veq, Cai:2021wzd, Inomata:2021tpx}. 
In contrast to the narrow case discussed above, the integrand in \eqref{H0_logst} is no longer concentrated around the peak. Reference \cite{Pi:2020otn} finds that the function $R$ in Eq.~\eqref{Rlogst} in the integrand behaves differently for $u \gtrsim 1$, $|u| \sim \mathcal{O}(1)$ and $u \lesssim -1$. Hence, Ref. \cite{Pi:2020otn} decomposes the integral into these three different domains, evaluates each separately, and adds up all the contributions at the end to obtain a formula; it turns out that the approximated results obtained by Ref. \cite{Pi:2020otn} are reasonably good compared with the numerical results.

For our case, we need to consider the behaviors of the combinations $\tilde{Q}_{\ell}^\text{LN} R$ in the integral region. 
Their analytic expressions are quite complicated, so it is not straightforward to get the semianalytic results as Ref. \cite{Pi:2020otn} did, and we leave this to the follow-up work. Here, we resort to the numerical method to calculate the integrals in Eq. \eqref{H0_logst} 
with different broad peaks $\Delta = 1, 10$ and the anisotropic coefficients $A_2 = \pm 0.2$, which are shown in the bottom-left and bottom-right panels of Fig. \ref{fig}. For comparison, we also plot the case $A_2 = -0.4$ for $\Omega_{2}^{\text{LN}}$.

As we expect, the multipole moments of SIGW energy spectra are extended as the width of the source increases.
Observing the bottom-left and bottom-right panels of Fig. \ref{fig}, we see that the zero $\tilde{k}_{21}^{\text{LN}}$ for $A_2 = 0.2$ of the quadrupole moment for $\Delta=1$ shifts to the larger-$k$ regime (while $\tilde{k}_{21}^{\text{LN}}$ for $A_2 = -0.2$ shifts to the smaller-$k$ regime) compared with the delta-function-like case shown in the top-right panel of Fig. \ref{fig}. 
We anticipate that the zeros move to the right as we increase the width, as we finally get almost flat spectra in the bottom-right panel in the figure. 
Also, the zeros of $\Omega_{2}^{\text{LN}}$ depend on the value of $A_2$ (see the cyan and black curves), since $\Omega_{2}^{\text{LN}}$ contains both the linear and quadratic terms in $A_2$, see Eq. \eqref{Q2}.
However, this behavior will not be explained straightforwardly based on the delta-function case. The log-normal case is not a simple superposition of delta-function sources, since Fourier mode coupling appears at second order. We should note that the zeros also depend on the magnitude of $A_2$.

\section{Conclusions} \label{sec:conclusion}

With the advent of the space-based GW experiments (e.g., LISA, DECIGO, TianQin, and Taiji), observations of SGWBs will play an irreplaceable role in telling us valuable and unique information about the early Universe.
In particular, SIGWs can be a probe of large scalar perturbations at tiny scales inaccessible by the CMB anisotropies.
We have very little information about such an extremely tiny scale so far, and even fundamental assumptions about statistical symmetry of the perturbations are not guaranteed at these scales.
This work considered the possibility of probing the statistical isotropy of the primordial density perturbations by using the SIGWs.

First, we reviewed the dynamics of the SIGWs and the forms of the source term. 
Then, we derived 
generic expressions of the multipole moments of the SIGW energy spectrum from an anisotropic scalar power spectrum. 
We showed that the monopole, quadrupole, and $\ell=4$ moments arise due to the quadrupole anisotropy in the scalar power spectrum.
This conclusion is independent of the shape of the anisotropic scalar power spectrum.  
Next, we considered two examples of the scalar power spectrum: the delta-function-like, and log-normal spectra during the radiation-dominated epoch. For the former case, we derived the analytic expressions for the multipole moments of the differential SIGW energy spectra. The monopole moment differs from the statistically isotropic case, and there exists a dip in the high-$k$ tail, which is a unique feature of the monopole moment, as shown in Fig. \ref{fig}. We showed the peculiar scale dependence of the multipole moments in the subhorizon scales, but the infrared behaviors are the same as with the isotropic one.
We considered the narrow $\Delta \ll 1$ and broad peak $\Delta \gtrsim 1$ separately for the latter log-normal spectrum. An isotropic narrow peak leads to a spectrum similar to the isotropic delta-function-like source but behaves as $k^3$ in the infrared tail, which also applies to the multipole moments in the anisotropic case.
For the broad peak, we perform the numerical calculations and get the moments of the differential SIGW energy spectra for $\Delta = 1, 10$ and $A_2 = \pm 0.2$, shown in Fig. \ref{fig}. As we expect, the SIGW spectra become broader for larger $\Delta$.

This work considered SIGWs from the anisotropic scalar power spectrum phenomenologically, whose amplitude is controlled by the size of scalar perturbations. Hence, this type of SIGW will be much more observationally interesting when one considers the PBH formation in the early Universe.
However, to our knowledge, PBH formation has not been discussed in the presence of statistical anisotropy.
We may consider additional vector fields in the existing PBH models, but it would be more interesting if the vector field itself can source a large scalar power spectrum.
In an anisotropic inflation scenario, $g_2$ in Eq.~\eqref{gstar} is somewhat a free parameter controlled by a gauge kinetic function in supergravity action~\cite{Soda:2012zm}.
If $g_0\ll g_2$ is realized at some small scales, $\mathcal P_\zeta(p)$ will be enhanced by $g_2$.
We will consider the possibility of such a PBH formation scenario in the follow-up work.

\section*{Acknowledgments}

We are grateful to Yudong Luo and Xi Tong for valuable discussions. C. C. especially thanks Yudong Luo for his help.
This work is supported in part by the National Key R\&D Program of China (No. 2021YFC2203100).
The authors are supported by the Jockey Club Institute for Advanced Study at The Hong Kong University of Science and Technology.
C. C. is grateful to Lei for her support.

\appendix

\begin{widetext}

\section{The derivation of the anisotropic spectrum of SIGWs} 

\label{app}

Starting from the EoM [Eq. \eqref{eom_hk}], the power spectrum of SIGWs from a statistically anisotropic source can be derived by using the Green's function solution of Eq. \eqref{eom_hk}, similar to the calculations of isotropic SIGWs \cite{Baumann:2007zm, Bartolo:2018rku, Cai:2019jah}, 
\be \label{ph_kpA}
\mathcal{P}_h^{\hat{ \mathbf{d} }, \lambda s}(\tau, \mathbf{k})
=
{k^3 \over \pi}
\int\mathrm{d}^3\mathbf{p}
\mathbf{e}^\lambda(\mathbf{k},\mathbf{p})
\mathbf{e}^s(\mathbf{k},\mathbf{p})
F(\tau, p, |\mathbf k-\mathbf p|)
{ \mathcal{P}_\zeta^{\hat{ \mathbf{d} }}(\mathbf p) \over p^3} 
{ \mathcal{P}_\zeta^{\hat{ \mathbf{d} }}(\mathbf{k} - \mathbf{p}) \over |\mathbf{k} - \mathbf{p}|^3 } ~,
\ee
where the source kernel $F(\tau, p, |\mathbf k-\mathbf p|)$ is shown in Eq. \eqref{source_kernel}. 
We consider angular dependence in the linear scalar power spectrum in Eq.~\eqref{ph_kpA}.
The preferred direction $\hat{\mathbf d}$ introduces the nontrivial azimuthal dependence for $\mathbf p$, which differs from the standard isotropic calculation.  
Without loss of generality, we may take a coordinate system where $\mathbf{k} = (0,0,k)$, $\mathbf{p} = p (\sin\theta \cos\varphi, \sin\theta \sin\varphi, \cos\theta)$ and $\hat{ \mathbf{d} } = (\sin\alpha \cos\beta, \sin\alpha \sin\beta, \cos\alpha)$, so that
\bl
\hat{ \mathbf{d} } \cdot \hat{\mathbf{p}}
=&
\cos\alpha \cos\theta
+ \cos(\varphi - \beta) \sin\theta \sin\alpha ,
\\
\hat{ \mathbf{d} } \cdot (\widehat{\mathbf{k} - \mathbf{p}}) 
=&
{ (k - p \cos\theta) \cos\alpha - p \cos(\varphi - \beta) \sin\alpha \sin\theta \over \sqrt{k^2 + p^2 - 2 k p \cos\theta} } ~.
\el
We change the variables as $x = |\mathbf{k} - \mathbf{p}| / k$, $y = p/k$ and $z = k\tau$. 
As we have $x=x(p,\theta)$ and $y=y(p)$, we can integrate $\varphi$ independently from $x$ and $y$ in Eq.~\eqref{ph_kpA}.
Then we get
\begin{align}
\label{tenkai}
\begin{split}
    \sum_{\lambda = +,\times} \mathcal{P}_h^{\lambda \lambda}(\tau, \mathbf{k})
=&   
P_0( \hat{ \mathbf{d} }\cdot \hat{\mathbf{k}} )  \int_{xy} A_{2}(kx) A_{2}(ky)Q_{0xy}(x,y) \\
&
- 5 P_2( \hat{ \mathbf{d} } \cdot \hat{\mathbf{k}} )  \int_{xy} \left[
A_{2}(kx) Q_{2x}(x,y)
+
A_{2}(ky) Q_{2y}(x,y)
+
A_{2}(kx) A_{2}(ky)
Q_{2xy}(x,y)
\right]\\
&+9 P_4( \hat{ \mathbf{d} } \cdot \hat{\mathbf{k}} )\int_{xy} 
A_{2}(kx) A_{2}(ky)
Q_{4xy}(x,y) ~,
\end{split}
\end{align}
where $\hat{ \mathbf{d} } \cdot \hat{\mathbf{k}}= \cos \alpha$ and we define
\be 
\int_{xy}
\equiv
\int_0^{\infty} dy
\int_{|1 - y|}^{1 + y} dx
\Big[ { 4 y^2 - (1 + y^2 - x^2 )^2 \over 4 x y} \Big]^2
F(z, x, y)
\mathcal{P}_\zeta(kx)
\mathcal{P}_\zeta(ky) ~. 
\ee
We also introduce
\bl \label{Q0xy}
Q_{0xy}(x,y)
=&
{160 \over 81} \l( { 5 + 3 \omega \over 6+6\omega} \r)^4
{ 1 \over x^2 y^2 }
\Big[
3 x^{4}
+ 2 x^2 (y^2-3)
+ 3 (y^2-1)^2
\Big] ~,
\\ \label{Q2x}
Q_{2x}(x,y)
=&
{32 \over 81 } \l( { 5 + 3 \omega \over 6+6\omega} \r)^4
{ 1 \over x^2 y^2 }
\Big[ 3 x^4 y^2 + 2 x^2 y^2 (1 - 3 y^2)	+ 3 y^2 (y^2 -1)^2 \Big] ~,
\\ \label{Q2y}
Q_{2y}(x,y)
=&
{32 \over 81 } \l( { 5 + 3 \omega \over 6+6\omega} \r)^4
{ 1 \over x^2 y^2 }
\Big[ 3 x^2 (1 + y^2 - x^2)^2  - 4 x^2 y^2 \Big] ~,
\\ \label{Q2xy}
Q_{2xy}(x,y)
=&
- {160 \over 567 } \l( { 5 + 3 \omega \over 6+6\omega} \r)^4
{ 1 \over x^2 y^2 }
\Big[
3 x^{6}
- 3 x^4 ( y^2 + 1)
+ 3 ( y^2 - 1)^2 ( y^2 + 1)
- x^2 (3 + 2 y^2 + 3 y^4 )
\Big] ~,
\\ \label{Q4xy}
Q_{4xy}(x,y)
=&
{20 \over 567 } \l( { 5 + 3 \omega \over 6+6\omega} \r)^4
{ 1 \over x^2 y^2 }
\Big[
35 x^8
- 20 x^6 (3 + 7 y^2)
+ 6 x^4 (3 + 10 y^2 + 35 y^4)
 \nn
\\& \hspace{100pt}
+ 4 x^2 (1 + 3 y^2 + 15 y^4 - 35 y^6)
+ ( y^2 -1)^2 (3 + 10 y^2 + 35 y^4)
\Big] ~.
\el
In the above derivations, we use the relation $[P_2(x)]^2 = {18 \over 35} P_4(x) + {2 \over 7} P_2(x) + {1 \over 5} P_0(x)$ for the Legendre polynomials.

We apply the above generic formulas to the RD epoch, where $\omega =1/3$ and $\mathcal{H} = 1/\tau$. The Green's function of Eq. \eqref{eom_hk} is given by
\be \label{rd_green}
g_{k}(\tau, \tau_1)
=
\Theta(\tau - \tau_1)
{ \tau_1 \sin[k (\tau - \tau_1)] \over k \tau } ~,
\ee
and the source function $f_\text{RD}(z_1, x, y)$ in Eq. \eqref{source_kernel} for the RD epoch is calculated as
\be \label{source_fxy_rd}
\begin{aligned}
	f_\text{RD}(z_1, x, y)
	=&
	{27 \over x^3 y^3 z^3} 
	\Big[
	18 x y z^2 \cos{x z \over \sqrt{3}} \cos{y z \over \sqrt{3}}
	+ [ 54 - 6 (x^2 + y^2) z^2 + x^2 y^2 z^4 ] \sin{x z \over \sqrt{3}} \sin{y z \over \sqrt{3}}
	\\& \quad
	+ 2 \sqrt{3} y z ( x^2 z^2 - 9) \sin{x z \over \sqrt{3}} \cos{y z \over \sqrt{3}}
	+ 2 \sqrt{3} x z ( y^2 z^2 - 9) \sin{y z \over \sqrt{3}} \cos{x z \over \sqrt{3}}
	\Big] ~,
\end{aligned}
\ee 
which is equal to $3$ at $z=0$ and decays as $\sim z^{-2}$ at large $z$, so that the source term during the RD epoch quickly decays, and a large amount of SIGWs are mainly produced at the early stage of horizon entry. Since we observe SIGWs at the present epoch-i.e., $\tau \rightarrow \infty$ or $z \gg 1$-in this limit and taking the time average, we find~\cite{Kohri:2018awv}
\be
\overline{ I_{\text{RD}}^2(z \rightarrow \infty, x, y) }
=
{1 \over 2} \Big( {27 (x^2 + y^2 -3) \over 16 x^3 y^3 z} \Big)^2
\Big[
\Big(
- 4 x y
+ (x^2 + y^2 -3) \ln\Big| {3 - (x+y)^2 \over 3 - (x-y)^2 } \Big|
\Big)^2
+ \pi^2 (x^2 + y^2 -3)^2 \Theta(x + y - \sqrt{3})
\Big] ~.
\ee
With the above preparations, substituting the above expressions into Eq. \eqref{H0}, and using the relation \eqref{omega_Ph}, we can obtain the semianalytic expressions of the differential energy spectrum $\Omega_{\text{GW}}^{\hat{ \mathbf{d} }}(\tau, \mathbf{k})$ of SIGWs and the multipole expansion [Eq. \eqref{omega_expansion}] when the isotropic power spectrum of the curvature perturbations $\mathcal{P}_\zeta(k)$ is given. The functions $Q_{0xy}(x,y)$, $Q_{2x}(x,y)$, $Q_{2y}(x,y)$, $Q_{2xy}(x,y)$, $Q_{4xy}(x,y)$ at the RD epoch are calculated by taking $\omega =1/3$ in Eqs. \eqref{Q0xy}-\eqref{Q4xy}.

\end{widetext}

\bibliographystyle{apsrev4-1}
\bibliography{igw}
\end{document}